\documentclass[a4paper,twocolumn,showpacs]{revtex4}
\usepackage{amsmath,amsfonts,amssymb}
\usepackage{psfrag}
\usepackage{graphicx}
\begin{document}

\title{Tunable Lyapunov exponent in inverse magnetic billiards}

\author{Zolt\'an V{\"o}r{\"o}s$^1$, Tam\'as Tasn\'adi$^2$, J{\'o}zsef Cserti$^1$, P{\'e}ter Pollner$^1$}
\affiliation{$^1$Department of Physics of Complex Systems}

\affiliation{$^2$Department of Solid State Physics, E{\"o}tv{\"o}s
University, H--1117 Budapest, P\'azm\'any P{\'e}ter s{\'e}t\'any 1/A, HUNGARY}

\begin{abstract}
The stability properties of the classical trajectories of
charged particles are investigated in a two dimensional stadium-shaped
inverse magnetic domain, where the magnetic field is zero inside the
stadium domain and constant outside. In the case of infinite magnetic
field the dynamics of the system is the same as in the Bunimovich billiard,
i.e., ergodic and mixing. However, for weaker magnetic fields the phase space
becomes mixed and the chaotic part gradually shrinks.  
The numerical measurements of the Lyapunov exponent (performed with a
novel method) and the integrable/chaotic phase space volume ratio show 
that both quantities can be smoothly tuned by varying the external 
magnetic field.
A possible experimental realization of the arrangement is also discussed.

\end{abstract}


\maketitle

In the past two decades, developments in nanotechnology have made 
it possible to electrostatically confine a two-dimensional electron
gas (2DEG) in high mobility heterostructures~\cite{Houten}. 
In these systems the
dynamics of the electrons is dominated by ballistic motion. 
Recently, a new perspective of the research of semiconductor systems
has been emerged by the application of spatially inhomogeneous
magnetic fields. 
The inhomogeneity of the magnetic field can be 
 realized experimentally either by varying the topography of the  
electron gas~\cite{topografia}, or 
using ferromagnetic materials~\cite{ferro},
or depositing a superconductor on top of the 2DEG~\cite{supra}.
Numerous theoretical works also show the increasing interest in the study
of electron motion in inhomogeneous magnetic field~\cite{elmelet}.

The aim of this Letter is to present   
a novel, experimentally realizable ballistic
2DEG system which exhibits a crossover between a well known, ergodic
and mixing billiard system (the Bunimovich stadium
billiard~\cite{Bun:79}), and a pathological integrable system, as the
applied magnetic field is changed. 
We suppose that the system is in the ballistic regime, like in many
other works (see e.g.~\cite{Houten,antidot}), and our treatment is 
purely classical.
Two characteristic quantities of the dynamics of this 
so-called {\em inverse magnetic billiard} are calculated 
numerically as a function of the  external magnetic field $\beta$: the
Lyapunov exponent $\lambda(\beta)$ (of the dominating chaotic
component), and the integrable/chaotic phase space volume 
ratio $\varrho(\beta)$.  The obtained numerical results show that both
quantities are smooth functions of the magnetic field which means that
the global dynamics of the system passes continuously from the integrable 
($\beta=0$) to the fully chaotic case ($\beta=\infty$).
As we shall see below, there is also a clearly visible correlated dependence 
between the variation of the quantities $\lambda(\beta)$ and 
$\varrho(\beta)$.   
These results, i.e., the fact that {\em the degree of chaoticity can 
smoothly be tuned} by the external magnetic field, may motivate 
the experimental realization and study of our presently proposed system. 
Kosztin et al.\ have made similar investigations and observations in Andreev
billiard systems~\cite{Kosztin}.

\begin{figure}[t]
\psfrag{r=1}{$r=1$}
\psfrag{a=2}{$a=2$}
\psfrag{R}{$R_c$}
\psfrag{c}{}
\psfrag{X}{$\mu$}
\psfrag{A}{$A$}
\includegraphics[scale=0.3]{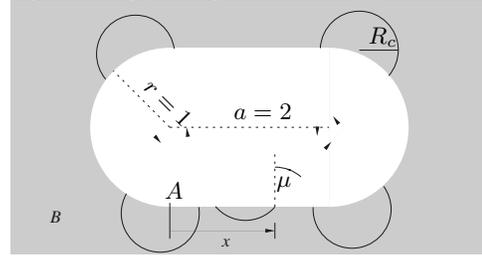}
\caption{\label{f:traj} The trajectories of a charged particle in the
inverse magnetic billiard. The cyclotron radius is $R_c=1/\beta =1/2$,
in dimensionless units.}
\end{figure}

More specifically, the system we suggest is a 2DEG in an
inhomogeneous magnetic field applied perpendicularly to the
system.
The magnetic field is considered to be zero inside
a stadium-shaped region and constant $\beta$ outside.
This arrangement can be realized experimentally by depositing a
stadium-shaped superconductor patch on the top of a 2DEG and applying
an external homogeneous magnetic field perpendicular to this structure. 
The magnetic field is excluded from the region covered by the
superconductor, due to the Meissner effect. 
A part of a typical classical
trajectory is depicted in Fig~\ref{f:traj}, for an intermediate value 
of the magnetic field $\beta= 2$.
The trajectories in the configuration space are straight segments inside
the stadium, and circular arcs of cyclotron radius 
$R_c=\frac{1}{\beta}$ out of this domain. 
(We assume, for simplicity, that the particle has unit mass,
charge and speed.) At the boundary of the domain the two pieces of the
trajectory join tangentially.  As the magnetic
field tends to infinity, $\beta \to \infty$, the electrons spend less and
less time outside the stadium, and it is also easy to see
that in the limiting case their motion is
described by an elastic reflection from the wall. For this reason we call our
system {\it inverse magnetic billiard}, although in the case of finite
field no real scatterings take place at the boundaries.

\begin{widetext}
\begin{figure*}
\psfrag{cos}{$\sin \mu$}
\psfrag{x}{$x$}
\includegraphics[scale=0.45]{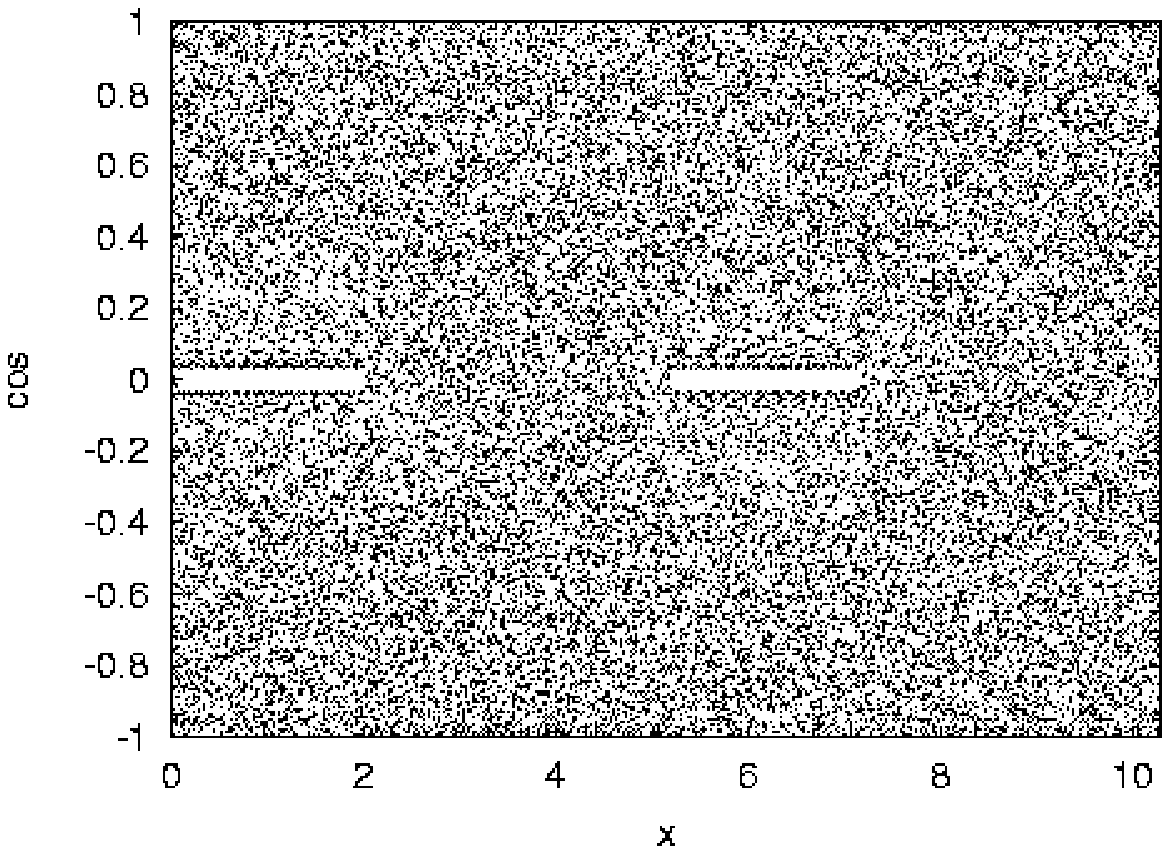}
\includegraphics[scale=0.45]{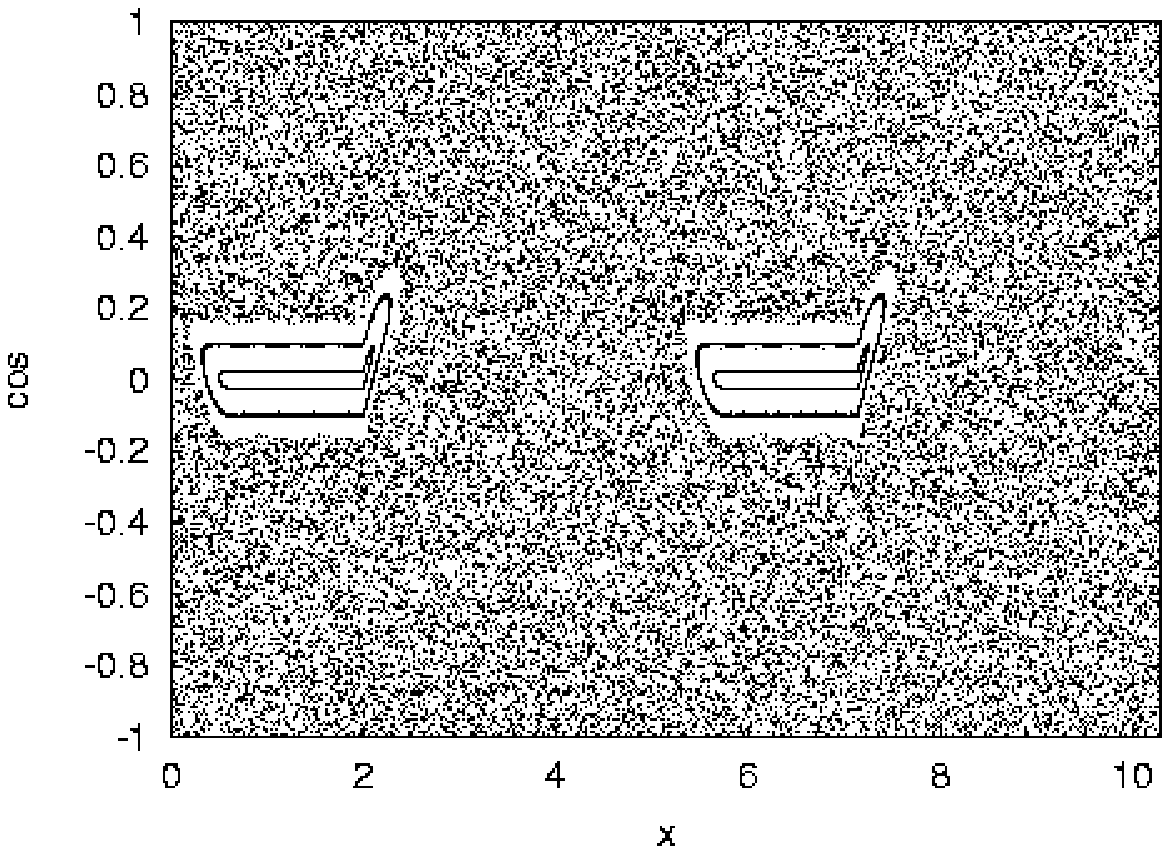}
\includegraphics[scale=0.45]{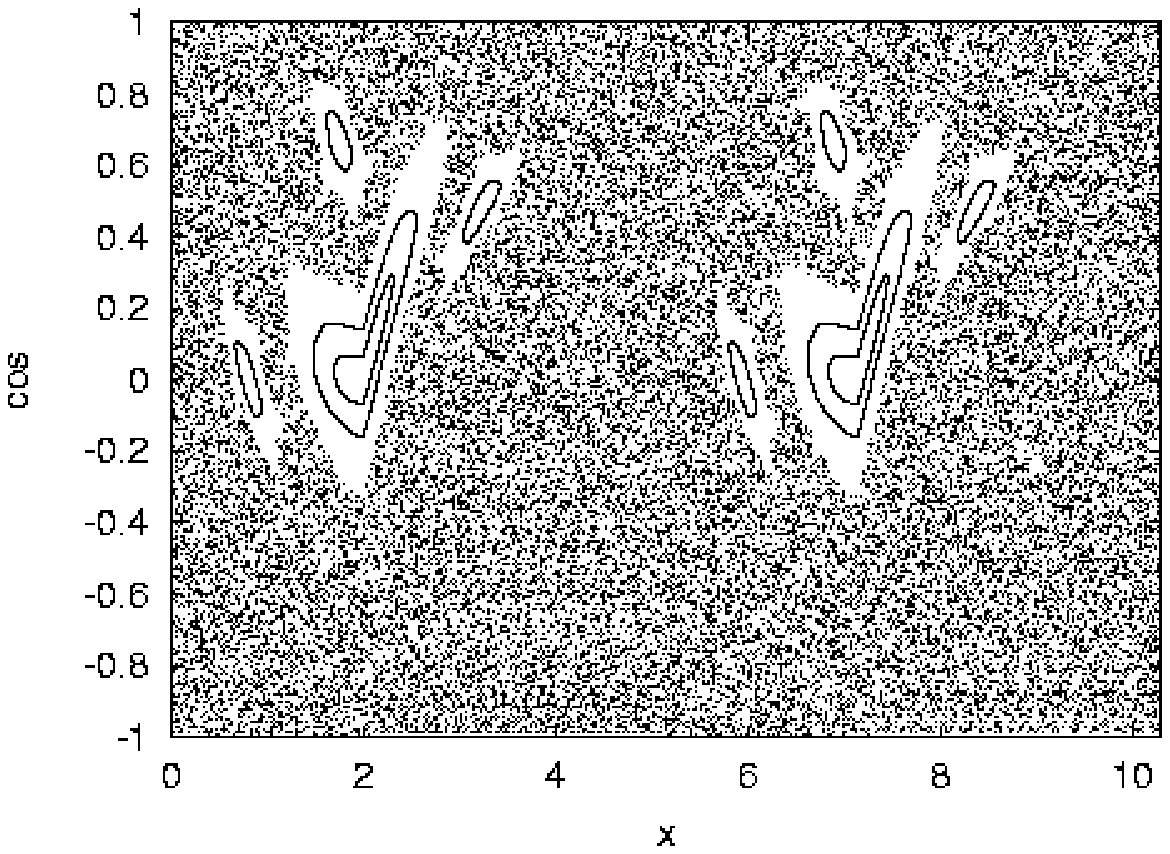} \\
{\qquad (a)}\hfill {\qquad (b)}\hfill {\qquad (c)}\hspace*{\fill}
\caption{\label{f:Poinc} The Poincar{\'e} section of the phase
space. The points in the dominating chaotic region
were obtained by 50000 iterations of a single
trajectory, while for depicting the islands corresponding to the
integrable regions, a few different initial conditions were used. The
values of the cyclotron radii are $R_c =0.05$, $R_c =0.3$, $R_c =1$,
respectively.}
\end{figure*}
\end{widetext}

According to the result of Bunimovich \cite{Bun:79}, the
stadium-shaped inverse magnetic billiard system is ergodic and mixing in
the $\beta=\infty$ case, but as the magnetic field is decreased, the
dynamics becomes partially integrable and gradually more and more phase
space volume is occupied by the KAM tori (mixed phase space).
This phenomenon can clearly be observed on the Poincar{\'e} sections
(see Fig.~\ref{f:Poinc}) made for different
magnetic field values. The individual points in the Poincar{\'e} sections
are plotted each time the particle enters the zero magnetic field
region and crosses the boundary of the stadium. The $x$ coordinate of
the points ($0\le x < 4+2\pi$) gives the position of the crossing,
measured in anti-clockwise direction from the point $A$ along 
the perimeter of the stadium, while the $y$ coordinate of the 
points ($-1 \le y \le 1$) denotes the sine of
the angle $\mu$ representing the direction of the trajectory, relative
to the normal of the boundary  (see Fig.~\ref{f:traj}).
It is well-known that in this parameter space the  Poincar{\'e} map is
area preserving~\cite{arepres_tangent_cikkek}.

It is evident from Fig.~\ref{f:Poinc} that for high magnetic fields the
system is (almost) completely chaotic but with decreasing magnetic field,
the volume of the integrable regions gradually increases. As we have seen
before, for $\beta=\infty$ the system is identical to the Bunimovich
billiard, however, in the $\beta \to 0$ limit the system becomes
pathological in the sense that the cyclotron radius tends to infinity, so
the electron returns to the stadium domain after longer and longer time
intervals.

\begin{figure}  
\psfrag{radius}{$R_c$}
\psfrag{ratio}{\small $\varrho(R_c)$}
\includegraphics[scale=0.5]{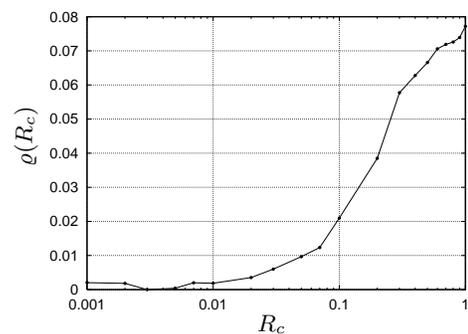}
\caption{\label{f:area} The integrable/chaotic phase space volume
ratio as a function of $R_c =1/\beta$.}
\end{figure}

In order to quantitatively characterize this change of the phase space
portrait we have  numerically investigated the integrable/chaotic phase 
space volume ratio $\varrho$  as a function of the cyclotron radius 
$R_c =1/\beta$ (i.e., the inverse magnetic field), and the results are 
shown in Fig.~\ref{f:area}. 
The function $\varrho(R_c)$, measured by the box-counting method with 
a grid of $250\times 250$ rectangular sites, is smooth, 
and its behavior is characteristically different for
higher and lower magnetic fields. For cyclotron radii less than $R_1
\approx 0.01$ (i.e., for magnetic fields larger than $\beta_1 \approx 100$)
the system is dominantly chaotic, the area of the integrable phase space
regions is practically negligible (see also Fig.~\ref{f:Poinc}.a).
For cyclotron radii larger than $R_2 \approx 0.3$, however, the chaotic part 
increases on the Poincar{\'e} section (see
also Fig.~\ref{f:Poinc}.c). Between these two extremities, i.e., for
cyclotron radii comparable to the characteristic size of the billiard, the
phase space of the system is definitely mixed (Fig.~\ref{f:Poinc}.b)
with integrable islands of considerable area.

Although the volume of the chaotic bands inside the integrable
islands (ignored in our treatment) is nonzero in principle,
the numerical simulations demonstrate (see Fig~\ref{f:Poinc}) that their
contribution to the chaotic phase space volume is negligible for this
system.

\begin{figure}   
\psfrag{radius}{$R_c$}
\psfrag{exp}{$\lambda(R_c)$}
\includegraphics[scale=0.5]{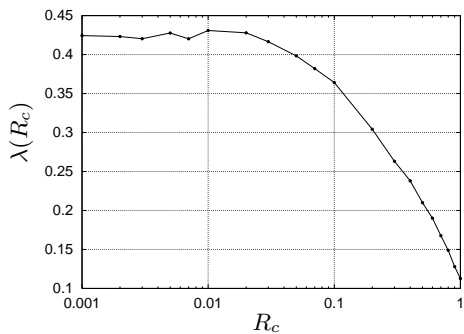}
\caption{\label{f:Lyap} Lyaponov exponent as a function of
$R_c= 1/\beta$.}
\end{figure}

Since the positivity of the Lyapunov exponent $\lambda(R_c)$ is one of  
the most characteristic features of chaotic systems, 
we have also numerically computed $\lambda(R_c)$
of the dominating chaotic component as a function of the cyclotron 
radius $R_c$ (see Fig.~\ref{f:Lyap}). 

The obtained function $\lambda(R_c)$ is again smooth, 
as $\varrho(R_c)$. It is also
clearly visible that the numerical value of the Lyapunov exponent strongly
correlates with the integrable phase space ratio $\varrho(R_c)$ measured
previously. For weak magnetic fields 
(if $\beta \lessapprox \beta_2 \approx 2$) the Lyapunov exponent is also
small, but as the magnetic field grows, the value of $\lambda$ increases,
too, and for strong fields (if $\beta \gtrapprox \beta_1 \approx 100$) 
it saturates at the value
$\lambda_{\infty} \approx 0.43$, which agrees well with the
Lyapunov exponent of the ordinary Bunimovich
billiard~\cite{Stadion_ljap}. 

In order to measure the Lyapunov exponent, we have investigated the
infinitesimal variations of the trajectories with the method of
Jacobi fields, which was originally developed for the stability analysis of
the geodetic flow on curved Riemannian manifolds \cite{KoNo:69}. The method
has successfully been applied to magnetic billiard systems on planar
\cite{TT:97} as well as curved surfaces \cite{TT:96,TT:98}. The main idea
of the method is to study the evolution of the so-called Jacobi fields
along a particular trajectory in the configuration space, which describe the
infinitesimal variations of the trajectory. This technique is essentially
the same as the method using the tangent map
\cite{arepres_tangent_cikkek}, but our approach is more transparent. 
The basic technical novelty is that in our investigations 
the coordinates describing the infinitesimal variations are chosen 
in a more natural way: they are
related to the unvaried trajectory itself, and not to the somewhat
artificial parameters of the space of the Poincar{\'e} section. As a result,
the stability matrices (i.e., the tangent maps) have a much simpler form.

In more details, let $\gamma_0(t)$ denote the trajectory in the
configuration space $\cal M$, whose stability properties we intend to
investigate, and let $\gamma_\varepsilon (t)$ be a one-parameter family of
varied trajectories around the unvaried one $\gamma_0$, i.e., for all
$\varepsilon \in (-\varepsilon_0, \varepsilon_0)$, $\varepsilon_0 >0$ the
curve $\gamma_\varepsilon$ is a real trajectory in the configuration space,
$\gamma_{\varepsilon =0} =\gamma_0$, and the map $\gamma : (-\varepsilon_0,
\varepsilon_0) \times {\mathbb R} \to {\cal M}$, $(\varepsilon,t) \mapsto
\gamma_{\varepsilon} (t)$ is everywhere continuous, and piecewise smooth.
(It is not smooth at the boundary of the billiard.) The {\it Jacobi field}
or {\it infinitesimal variation vector field} $V_{\gamma_0}$ corresponding
to the variation $\gamma_{\varepsilon}$ is the partial derivative 
$V_{\gamma_0} (t)=\left. \frac{\partial \gamma_{\varepsilon}
(t)}{\partial \varepsilon}\right|_{\varepsilon=0}$.

It can be shown that the Jacobi fields $V_{\gamma_0} (t)$ satisfy certain
second order differential equation, called {\it Jacobi equation}; it is
due to the fact that the varied curves $\gamma_{\varepsilon}$ are also real
trajectories \cite{KoNo:69,TT:96}. In two dimensional billiard systems we
found it convenient to fix the base vectors $\big\{ \Dot{\gamma}_0(t),
\Dot{\gamma}_0^{\perp} (t) \big\}$ of the coordinate system to the
investigated trajectory $\gamma_0 (t)$, in such a way that
$\Dot{\gamma}_0(t)$ is the (unit) vector tangential to the trajectory at
the time instant $t$, and $\Dot{\gamma}_0^{\perp} (t)$ is obtained from
$\Dot{\gamma}_0(t)$ by a rotation through $+90^{\circ}$. In this basis the
Jacobi field is written as $V_{\gamma_0} (t) =\xi(t) \cdot
\Dot{\gamma}_0(t) +\eta(t) \cdot \Dot{\gamma}_0^{\perp} (t)$, and for
characterizing a given infinitesimal variation the initial conditions
$\xi(t_0)$, $\eta(t_0)$, $\Dot{\xi}(t_0)$ and $\Dot{\eta}(t_0)$ have to be
given. (The real functions $\xi$ and $\eta$ are the coordinates of
the Jacobi field $V_{\gamma_0}$.)

The number of these initial data can further be reduced by two, if
we notice that {\it i)} the longitudinal variations $\xi(t)$ as well as
{\it ii)} the variations altering the speed (i.e., for which $\dot{\xi}
\ne 0$) are irrelevant in the present investigation, and they decouple from
the other coordinates, so they can be disregarded. (In the case {\it i)} the
Jacobi field is tangential to the unvaried trajectory $\gamma_0$, thus the
varied curves are just time-shifts of the original one, while {\it ii)}
means that we restrict the attention to a constant energy shell of the
phase space, as it is usual in Hamiltonian systems.)

In planar billiard systems it is an elementary geometric problem to find
the solutions of the Jacobi equation in terms of the transverse coordinates
$\eta (t)$ and $\Dot{\eta}(t)$ (see e.g. \cite{TT:97}).
Generally, the solution is given by a linear transformation $\Big[
\begin{smallmatrix} \eta' \\ \Dot{\eta}' \end{smallmatrix} \Big]=
\mathbf{L} \Big[ \begin{smallmatrix} \eta \\ \Dot{\eta} \end{smallmatrix}
\Big]$, where the matrix $\mathbf L$ has the following special forms for
the straight flight in zero magnetic field ($\mathbf P$), for the curved
flight in nonzero magnetic field ($\mathbf E$) and for the boundary
transition ($\mathbf T$) with magnetic field change $\Delta \beta$,
respectively:
\begin{subequations}\label{e:PET}
\begin{align}
\mathbf{P}(t) &= \begin{bmatrix} 1
& t \\ 0 & 1 \end{bmatrix},
\\
\mathbf{E}(t,\beta) &= \begin{bmatrix} \cos(\beta t) &
\frac{1}{\beta} \sin(\beta t) \\ -\beta \sin(\beta t) & \cos(\beta t)
\end{bmatrix},
\\
\mathbf{T}(\Delta \beta, \mu) &= \begin{bmatrix} 1 & 0 \\
\Delta \beta \tan \mu & 1 \end{bmatrix}.
\end{align}
\end{subequations}
Here $t$ is the time of flight (so $\beta t$ is the angle of flight),
$\beta$ denotes the magnetic field and $\mu$ is the angle of incidence at
the boundary, measured in the way shown in Fig~\ref{f:traj}. It is worth
noticing that all the three types of matrices are one-parameter subgroups
of $SL(2,{\mathbb R})$, i.e., of the group of two by two real
matrices with unit determinant. The matrices $\mathbf P$ and $\mathbf T$
are parabolic, while the transformations $\mathbf E$ are elliptic.

For investigating the long time stability of a given trajectory $\gamma_0$
the eigenvalues (or the trace) of the product matrix
\begin{equation}\label{e:TETP}
\dots ({\mathbf T}'_3{\mathbf E}_3{\mathbf T}_3{\mathbf P}_3)
({\mathbf T}'_2{\mathbf E}_2{\mathbf T}_2{\mathbf P}_2)
({\mathbf T}'_1{\mathbf E}_1{\mathbf T}_1{\mathbf P}_1)
\end{equation}
have to be calculated, where the individual matrices in the expression
describe, in reverse order, the stability of the corresponding segments of
the motion (in the billiard, through the boundary outwards, in the magnetic
field and back again into the billiard through the boundary). 
This group of four matrices corresponds to a cycle in 
the Poincar{\'e} sections of Fig.~\ref{f:Poinc}. 
(The matrices ${\mathbf T}$, ${\mathbf T}'$
correspond to the outward and inward passage through the boundary,
respectively.)

In our simulations the matrices \eqref{e:PET} and the product
\eqref{e:TETP} corresponding to about  25000 cycles were calculated 
explicitly, and the Lyapunov exponents, shown in Fig.~\ref{f:Lyap} 
were computed as the logarithm of the largest eigenvalue 
(practically, the trace) of the resulting matrix
divided by the total time of flight.

The fact that in the $\beta \to \infty$ limit the inverse magnetic billiard
gives back the dynamics of the normal billiard system with elastic walls
can be checked also in terms of the stability matrices. A bit lengthy but
straightforward calculation yields that if the billiard wall is a circle of
curvature $q$, then
\begin{equation}
\lim_{\beta \to \infty} \big( {\mathbf T}(-\beta, -\mu){\mathbf
E}(t, \beta){\mathbf T}(\beta, \mu) \big) =-
\begin{bmatrix} 1 & 0 \\ -\frac{2q}{\cos \mu} & 1 \end{bmatrix},
\end{equation}
which is the stability matrix corresponding to an elastic reflection on the
wall of curvature $q$ \cite{TT:96}, as it is expected. (The signs of the
arguments of $\mathbf T$ can be obtained by elementary geometric
considerations.)


We now turn to the discussion of the conditions of the experimental
realization of the inverse magnetic billiards using GaAs/AlGaAs
heterostructure. 
There are four characteristic lengths in the system: the Fermi
wavelength (typically $\lambda_{\rm{F}}=40$~nm~\cite{Houten}), 
the radius  $r$ of the stadium, the
cyclotron radius $R_c$ and the mean free path $l$ (which can be as
high as $10^4$~nm~\cite{Houten}).  
The classical ballistic motion of the electrons 
requires that $\lambda_{\rm{F}}\ll r, R_c \ll  l$. (The last condition
assures that the electron travels through several
Poincar{\'e} cycles without scattering on impurities.)   
Fig.~\ref{f:Lyap} shows that
the relevant values of the ratio $r/R_c$ are in the range of $0.01-1.0$.
The magnetic field  
can be as high as $2$ T without destroying
superconductivity. This implies that $R_c  \gtrapprox 50$~nm 
(using that the effective mass of electrons $m_{\rm{eff}}=0.067m_e$, 
where $m_e$ is the mass of the electron,
and $E_{\rm{F}} = 14 \,\,\mathrm{meV}$~\cite{Houten}).
Assuming that the size of a superconductor grain is about $r=1$~$\mu$m,    
the cyclotron radii are $50,\, 300,\, 1000$~nm corresponding to
data $R_c/r$ in Fig.~\ref{f:Poinc}. This implies that 
 parameter $\beta$ in  Fig.~\ref{f:Poinc} corresponds to
the experimental values of the magnetic field $ 2, 0.3, 0.2$~T, respectively.
It is clear that these experimental values do not perfectly fit the condition 
of the classical motion. The semiclassical or full quantum mechanical 
treatment of the problem can be an extension of our work. 

The advantage of our suggested setup in comparison with 
Andreev billiards (which is another proposed 
experimental setup for magnetically tunable chaoticity) 
is that in our system the electrons travel in a homogeneous
heterostructure without any scattering on the boundary of the stadium, 
whereas in the case of Andreev billiards the normal reflections may 
suppress the effect as discussed in Ref.~\onlinecite{Kosztin}. 

We remark that in a real experiment, the profile of the magnetic field 
cannot be approximated by a step function as we assumed before. 
However, the deviation of the magnetic field from the sharp profile 
can easily be included in classical calculations.   

In practice, one would measure the conductance or susceptibility, 
which should be sensible to the chaotic nature of the system tuned by 
magnetic field \cite{Marcus}.

One of us (J.\ Cs.) gratefully acknowledges very helpful discussions
with C. Lambert and A. Voros.
This work is supported in part by 
the Hungarian-British Intergovernmental Agreement on Cooperation in
Education, Culture, and Science and Technology, 
and the Hungarian  Science Foundation OTKA  TO34832 and D37788.


\end{document}